# Adaptive Conditional Bias-Penalized Kalman Filter for Improved Estimation of Extremes and Its Approximation for Reduced Computation

Haojing Shen, Haksu Lee, and Dong-Jun Seo

*Abstract*— **In many signal processing applications of Kalman filter (KF) and its variants and extensions, accurate estimation of extreme states is often of great importance. When the observations used are uncertain, however, KF suffers from conditional bias (CB) which results in consistent under- and overestimation of extremes in the right and left tails, respectively. Recently, CB-penalized KF, or CBPKF, has been developed to address CB. In this paper, we present an alternative formulation based on variance-inflated KF to reduce computation and algorithmic complexity, and describe adaptive implementation to improve unconditional performance. For theoretical basis and context, we also provide a complete self-contained description of CB-penalized Fisher-like estimation and CBPKF. The results from 1-dimensional synthetic experiments for a linear system with varying degrees of nonstationarity show that adaptive CBPKF reduces root mean square error at the extreme tail ends by 20 to 30% over KF while performing comparably to KF in the unconditional sense. The alternative formulation is found to approximate the original formulation very closely while reducing computing time to 1.5 to 3.5 times of that for KF depending on the dimensionality of the problem. Adaptive CBPKF hence offers a significant addition to the dynamic filtering methods for general application in signal processing when accurate estimation of extremes is of importance.**

*Index Terms*— **Adaptive filtering, Conditional bias, Extremes, Kalman filter**

## I. INTRODUCTION

Kalman filter (KF) and its variants and extensions are widely used to fuse observations with model predictions in a wide range of applications [1][2][3][4][5][6][7][8][9][10][11][12][13][14][15][16][17][18][19][20][21]. In geophysics and environmental science and engineering, often the main objective of signal processing is to improve estimation and prediction of states in their extremes rather than in normal ranges. In hydrologic forecasting, for example, accurate prediction of floods and droughts is far more important than that of streamflow and soil moisture in normal conditions. Because KF minimizes unconditional error variance, its solution tends to improve estimation near median where the state of the dynamic system resides most of the times while often leaving significant biases in the extremes. Such conditional biases (CB)[22]generally result in consistent under- and overestimation of the true states in the upper and lower tails of the distribution, respectively. To address CB, CB-penalized Fisher-like estimation and CB-penalized KF (CBPKF)[23][24]have recently been developed which jointly minimize error variance and expectation of the Type-II CB squared for improved estimation and prediction of extremes. The Type-II CB, defined as $E[\hat{X} | X = x] - x$, is associated with failure to detect the event where $x$ denotes the realization of $X$ where $X$, $\hat{X}$ and $\hat{x}$ denote the unknown truth, the estimate, and the realization of $\hat{X}$, respectively[25]. The original formulation of CBPKF, however, is computationally very expensive for high-dimensional problems. Also, whereas CBPKF improves performance in the tails, it deteriorates performance in the normal ranges. In this work, we approximate CBPKF with forecast error covariance-inflated KF, referred to hereafter as the variance-inflated KF (VIKF) formulation, as a computationally less expensive and algorithmically simpler alternative, and implement adaptive CBPKF to improve performance in the unconditional mean sense.

Elements of CB-penalized Fisher-like estimation has been described in the forms of CB-penalized indicator cokriging for fusion of predicted streamflow from multiple models and observed streamflow[26], CB-penalized kriging for spatial estimation[27] and rainfall estimation[28], and CB-penalized cokriging for fusion of radar rainfall and rain gauge data[29].The original formulation of CBPKF have been described in[24]and[23], respectively. Its ensemble extension, CB-penalized ensemble KF, or CEnKF, is described in [30] in the context of ensemble data assimilation for flood forecasting.

Submitted to IEEE Transactions on Signal Processing on Jan 8, 2019.
This work was supported by the National Science Foundation [CyberSEES-1442735], and the National Oceanic and Atmospheric Administration [NA16OAR4590232, NA17OAR4590174, NA17OAR4590184].

Haojing Shen is with Department of Civil Engineering, The University of Texas at Arlington, Box 19308, 416 Yates St. Suite 425, Arlington, TX 76019 (e-mail: Haojing.shen@mavs.uta.edu).

Haksu Lee is with Len Technologies, Oak Hill, VA 20171(e-mail: haksulee@hanmail.net).

Dong-Jun Seo is with Department of Civil Engineering, The University of Texas at Arlington, Box 19308, 416 Yates St. Suite 425, Arlington, TX 76019 (e-mail: djseo@uta.edu).

Whereas CBPKF was initially motivated for environmental and geophysical state estimation and prediction, it is broadly applicable to a wide range of applications for which improved performance in the extremes is desired. This paper is organized as follows. Sections II and III describe CB-penalized Fisher-like solution and CBPKF, respectively. Section IV describes approximation of CBPKF. Sections V describe the evaluation experiments and results, respectively. Section VI describes adaptive CBPKF. Section VII provides conclusions.

## II. CONDITIONAL BIAS-PENALIZED FISHER-LIKE SOLUTION

As in Fisher estimation[31], the estimator sought for CB-penalized Fisher-like estimation is $X^* = WZ$ where $X^*$ denotes the $(m \times 1)$ vector of the estimated states, $W$ denotes the $(m \times (n+m))$ weight matrix, and $Z$ denotes the $((n+m) \times 1)$ augmented observation vector. In the above, n denotes the number of observations, m denotes the number of state variables, and (n+m) represents the dimensionality of the augmented vector of the observations and the model-predicted states to be fused for estimation of the true state $X$. The purpose of augmentation is to relate directly to CBPKF in Section III without introducing additional notations. Throughout this paper, we use regular and bold letters to differentiate the non-augmented and augmented variables, respectively. The linear observation equation is given by:

$$Z = HX + V \quad (1)$$

where X denotes the $(m \times 1)$ vector of the true state with $E[X]=M_X$ and $Cov[X,X^T]=\Psi_{XX}$, $H$ denotes the $((n+m) \times m)$ augmented linear observation equation matrix, and $V$ denotes the $((n+m) \times 1)$ augmented zero-mean observation error vector with $Cov[V,V^T]=R$. Assuming independence between X and V, we write the Bayesian estimator[31] for $X$, or $X^*$, as:

$$X^* = M_X + W(Z - HM_x) \quad (2)$$

The error covariance matrix for $X^*$, $E[(X-X^*)(X-X^*)^T]$, is given by:

$$\Sigma_{EV} = (I - WH)\Psi_{XX}(I - WH)^T + WRW^T \quad (3)$$

With (2), we may write Type-II CB as:

$$X - E[X^*|X] = (X - M_X) - WE[(Z - HM_x)|X] \quad (4)$$

The observation equation for $Z$ is obtained by inverting (1):

$$X = GZ - GV \quad (5)$$

The (mx(n+m)) matrix, $G$, in (5) is given by:

$$G = (U^T H)^{-1} U^T \quad (6)$$

where $U^T$ is some $(m \times (n+m))$ nonzero matrix. Using (5) and the identity, $\Psi_{ZZ} = H\Psi_{XX}H^T + R$, we may write the Bayesian estimate for E[Z|X] in (4) as:

$$\hat{E}[Z|X] = HM_X + C(X - M_X) \quad (7)$$

where

$$C = (H\Psi_{XX}H^T + R)G^T[G(H\Psi_{XX}H^T + 2R)G^T]^{-1} \quad (8)$$

Equations (7) and (8) state that the Bayesian estimate of $Z$ given X is given by $HX$ if the a priori state error covariance $\Psi_{XX}$ is noninformative or there are no observation errors, but by the average of the a priori mean $M_X$ and the observed true state X if the a priori $\Psi_{XX}$ is perfectly informative or observations are information-less.

With (4), we may write the quadratic penalty due to Type-II CB as:

$$\Sigma_{CB} = E[(X - E_{X^*}[X^*|X])(X - E_{X^*}[X^*|X])^T] = (I - WC)\Psi_{XX}(I - WC)^T \quad (9)$$

where I denotes the $(m \times m)$ identity matrix. Combining $\Sigma_{EV}$ in (3) and $\Sigma_{CB}$ in (9), we have the apparent error covariance, $\Sigma_a$, which reflects both the error covariance and Type-II CB:

$$\Sigma_a = (I - WH)\Psi_{XX}(I - WH)^T + WRW^T + \alpha(I - WC)\Psi_{XX}(I - WC)^T \quad (10)$$

where α denotes the scaler weight given to the CB penalty term. Minimizing (10) with respect to $W$, or by direct analogy with the Bayesian solution[31], we have:

$$W = \Psi_{XX}\hat{H}^T[\hat{H}\Psi_{XX}\hat{H}^T + \Lambda]^{-1} \quad (11)$$

The modified structure matrix $\hat{H}^T$ and observation error covariance matrix $\Lambda$ in (11) are given by:

$$\hat{H}^T = H^T + \alpha C^T \quad (12)$$

$$\Lambda = R + \alpha(1-\alpha)C\Psi_{XX}C^T - \alpha H\Psi_{XX}C^T - \alpha C\Psi_{XX}H^T \quad (13)$$

Using (11) and the matrix inversion lemma[32], we have for $\Sigma_a$ and $X^*$ in (10) and (2), respectively:

$$\Sigma_a = \alpha\Psi_{XX} + [\hat{H}\Lambda^{-1}\hat{H}^T + \Psi_{XX}^{-1}]^{-1} \quad (14)$$

$$X^* = [\hat{H}^T\Lambda^{-1}\hat{H} + \Psi_{XX}^{-1}]^{-1}\{\hat{H}^T\Lambda^{-1}Z + \Psi_{XX}^{-1}M_X\} + \Delta \quad (15)$$

where $\Delta = \alpha\Psi_{XX}\hat{H}^T[\hat{H}\Psi_{XX}\hat{H}^T + \Lambda]^{-1}CM_X$. To render the above Bayesian solution to a Fisher-like solution, we assume no a priori information in X and let $\Psi_{XX}^{-1}$ in the brackets in (14) and (15) vanish:

$$\Sigma_a = B[\hat{H}\Lambda^{-1}\hat{H}^T]^{-1} \quad (16)$$

$$X^* = [\hat{H}^T\Lambda^{-1}\hat{H}]^{-1}\hat{H}^T\Lambda^{-1}Z + \Delta \quad (17)$$

where the scaling matrix B is given by $B = \alpha\Psi_{XX}\hat{H}^T\Lambda^{-1}\hat{H} + I$. To obtain the estimator of the form, $X^* = WZ$, we impose the unbiasedness condition, $E[X^*] = X$, or equivalently, $WH = I$. The above condition is satisfied by replacing $[\hat{H}^T\Lambda^{-1}\hat{H}]^{-1}$ with $[\hat{H}^T\Lambda^{-1}H]^{-1}$ and dropping $\Delta$ in (17):

$$\Sigma_a = B[\hat{H}\Lambda^{-1}H^T]^{-1} \quad (18)$$

$$X^* = [\hat{H}^T\Lambda^{-1}H]^{-1}\hat{H}^T\Lambda^{-1}Z \quad (19)$$

Finally, we obtain from (3) the error covariance, $\Sigma_{EV}$, associated with $X^*$ in (19):

$$\Sigma_{EV} = WRW^T = [\hat{H}^T\Lambda^{-1}H]^{-1}\hat{H}^T\Lambda^{-1}R\Lambda^{-1}\hat{H}[\hat{H}^T\Lambda^{-1}H]^{-1} \quad (20)$$

Note that, if α=0, we have $\hat{H}^T = H$ and $\Lambda = R$, and hence the

CB-penalized Fisher-like solution, (19) and (20), is reduced to the Fisher solution[31].

### III. CONDITIONAL BIAS-PENALIZED KALMAN FILTER

CBPKF results directly from decomposing the augmented matrices and vectors in (19) and (20) as KF does from the Fisher solution[31]. The CBPKF solution, however, is not very simple because the modified observation error covariance matrix, $\Lambda$, is no longer diagonal. An important consideration in casting the CB-penalized Fisher-like solution into CBPKF is to recognize that CB arises from the error-in-variable effects associated with uncertain observations[33], and that the a priori state, represented by the dynamical model forecast, is not subject to CB. We therefore apply the CB penalty to the observations only, and reduce $\mathbf{C}$ in (8) to $\mathbf{C}^T = (C_{1,k}^T C_{2,k}^T) = (C_{1,k}^T\ 0)$. Separating the observation and dynamical model components in $\widehat{H}^T$ and $\Lambda$ via the matrix inversion lemma, we have:

$$\widehat{H}^T = (\widehat{H}_{1,k}^T\ I) \tag{21}$$

$$\Lambda = \begin{bmatrix} \Lambda_{11,k} & \Lambda_{12,k} \\ \Lambda_{21,k} & \Lambda_{22,k} \end{bmatrix} \tag{22}$$

where

$$\widehat{H}_{1,k}^T = H_k^T + \alpha C_{1,k}^T \tag{23}$$

$$\Lambda_{11,k} = R_k + \alpha(1-\alpha)C_{1,k}\Psi_{XX}C_{1,k}^T - \alpha H_k \Psi_{XX} C_{1,k}^T - \alpha C_{1,k}\Psi_{XX}H_k^T \tag{24}$$

$$\Lambda_{12,k} = -\alpha C_{1,k}\Psi_{XX} \tag{25}$$

$$\Lambda_{21,k} = \Lambda_{12,k}^T \tag{26}$$

$$\Lambda_{22,k} = \Sigma_{k|k-1} \tag{27}$$

In the above, $H_k$ denotes the (n×m) observation matrix, and $R_k$ denotes the (n×n) observation error covariance matrix. To evaluate the (m×n) matrix, $C_{1,k}$, it is necessary to specify $\mathbf{U}^T$ in (6). We use $\mathbf{U}^T = \mathbf{H}^T$ which ensures invertibility of $\mathbf{U}^T\mathbf{H}$, but other choices are possible. We then have for $C_{1,k}^T$:

$$C_{1,k} = [(H_k\Psi_{XX}H_k^T + R_k)G_{1,k} + H_k\Psi_{XX}G_{2,k}]L_k^{-1} \tag{28}$$

where

$$G_{2,k}^T = (H_k^T H_k + I)^{-1} \tag{29}$$

$$G_{1,k}^T = G_{2,k}^T H_k^T \tag{30}$$

$$L_k = G_{2,k}^T [H_k^T(H_k\Psi_{XX}H_k^T + 2R_k)H_k + H_k^T H_k \Psi_{XX} + \Psi_{XX} H_k^T H_k + \Psi_{XX} + 2\Sigma_{k|k-1}]G_{2,k} \tag{31}$$

Expanding $\mathbf{W}$ in (11) with $\Lambda^{-1} = \Gamma = \begin{bmatrix} \Gamma_{11,k} & \Gamma_{12,k} \\ \Gamma_{21,k} & \Gamma_{22,k} \end{bmatrix}$, we have;

$$\mathbf{W} = [\widehat{H}^T\Lambda^{-1}H]^{-1}\widehat{H}^T\Lambda^{-1} = (\varpi_{1,k}H_k + \varpi_{2,k})^{-1}(\varpi_{1,k}\varpi_{2,k}) \tag{32}$$

In (32), the (m×n) and (m×m) weight matrices for the observation and model prediction, $\omega_{1,k}$ and $\omega_{2,k}$, respectively, are given by:

$$\varpi_{1,k} = \widehat{H}_{1,k}^T\Gamma_{11,k} + \Gamma_{21,k} \tag{33}$$

$$\varpi_{2,k} = \widehat{H}_{1,k}^T\Gamma_{12,k} + \Gamma_{22,k} \tag{34}$$

where

$$\Gamma_{22,k} = [\Lambda_{22,k} - \Lambda_{21,k}\Lambda_{11,k}^{-1}\Lambda_{12,k}]^{-1} \tag{35}$$

$$\Gamma_{11,k} = \Lambda_{11,k}^{-1} + \Lambda_{11,k}^{-1}\Lambda_{12,k}\Gamma_{22,k}\Lambda_{21,k}\Lambda_{11,k}^{-1} \tag{36}$$

$$\Gamma_{12,k} = -\Lambda_{11,k}^{-1}\Lambda_{12,k}\Gamma_{22,k} \tag{37}$$

The apparent CBPKF error covariance, which reflects both $\Sigma_{EV}$ and $\Sigma_{CB}$, is given by (18) as:

$$\Sigma_{a,k|k} = \alpha\Sigma_{k|k-1} + [\varpi_{1,k}H_k + \varpi_{2,k}]^{-1} \tag{38}$$

The CBPKF error covariance, which reflects $\Sigma_{EV}$ only, is given by (20) as:

$$\Sigma_{k|k} = [\varpi_{1,k}H_k + \varpi_{2,k}]^{-1}(\varpi_{1,k}R_k\varpi_{1,k}^T + \varpi_{2,k}\Sigma_{k|k-1}\varpi_{2,k}^T)[\varpi_{1,k}H_k + \varpi_{2,k}]^{-1} \tag{39}$$

Because CBPKF minimizes $\Sigma_{a,k|k}$ rather than $\Sigma_{k|k}$, it is not guaranteed that (39) satisfies $\Sigma_{k|k} \leq \Sigma_{k|k-1}$ a priori. If the above condition is not met, it is necessary to reduce α and repeat the calculations. If α is reduced all the way to zero, CBPKF collapses to KF. The CBPKF estimate may be rewritten into a more familiar form:

$$\widehat{X}_{k|k} = [\varpi_{1,k}H_k + \varpi_{2,k}]^{-1}[\varpi_{1,k}Z_k + \varpi_{2,k}\widehat{X}_{k|k-1}] = \widehat{X}_{k|k-1} + K_k[Z_k - H_k\widehat{X}_{k|k-1}] \tag{40}$$

In (40), $Z_k$ denotes the (n×1) observation vector, and the (m×n) CB-penalized Kalman gain, $K_k$, is given by:

$$K_k = [\varpi_{1,k}H_k + \varpi_{2,k}]^{-1}\varpi_{1,k} \tag{41}$$

To operate the above as a sequential filter, it is necessary to prescribe $\Psi_{XX}$ and α. An obvious choice for $\Psi_{XX}$, i.e., the a priori error covariance of the state, is $\Sigma_{k|k-1}$. Specifying α requires some care. In general, a larger α improves accuracy over the tails but at the expense of increasing unconditional error. Too small an α may not effect large enough CB penalty in which case the CBPKF and KF solutions would differ little. Too large an α, on the other hand, may severely violate the $\Sigma_{k|k} \leq \Sigma_{k|k-1}$ condition in which case the filter may have to be iterated at additional computational expense with successively reduced $\alpha$. A reasonable strategy for reducing $\alpha$ is $\alpha_i = c\alpha_{i-1}, i = 1,2,3,...$, with $0 < c < 1$ where $\alpha_i$ denotes the value of α at the i-th iteration[24][30]. For high-dimensional problems, CBPKF can be computationally very expensive. Whereas KF requires solving an (m×n) linear system only once per updating or fusion cycle, CBPKF additionally requires solving two (m×m) linear systems (for $C_{1,k}$ and $\Gamma_{22}$), and an (n×n) system (for $\Lambda_{11}$), assuming that the structure of the observation equation does not change in time (in which case $G_{2,k}^T$ in (29) may be evaluated only once). To reduce computation, below we approximate CBPKF with KF by inflating the forecast error covariance.

## IV. VIKF APPROXIMATION OF CBPKF

The main idea behind this simplification is that, if the gain for the CB penalty, $C$, in (10) can be linearly approximated with $H$, the apparent error covariance $\Sigma_a$ becomes identical to $\Sigma_{EV}$ in (3) but with $\Psi_{XX}$ inflated by a factor of $1+\alpha$:

$$\Sigma_{(1+\alpha)} = (I - WH)(1+\alpha)\Psi_{XX}(I - WH)^T + WR_{(1+\alpha)}W^T \quad (42)$$

where $R_{(1+\alpha)} = \begin{bmatrix} R & 0 \\ 0 & (1+\alpha)\Psi_{XX} \end{bmatrix}$. The KF solution for (42) is identical to the standard KF solution but with $\Sigma_{k|k-1}$ replaced by $(1+\alpha)\Sigma_{k|k-1}$:

$$\hat{X}_{k|k} = [H_k^T R_k^{-1} H_k + \{(1+\alpha)\Sigma_{k|k-1}\}^{-1}]^{-1}[H_k^T R_k^{-1} Z_k + \{(1+\alpha)\Sigma_{k|k-1}\}^{-1}\hat{X}_{k|k-1}] \quad (43)$$

With $WH = I$ in (43) for the VIKF solution, we have $\Sigma_{(1+\alpha)} = WR_{(1+\alpha)}W^T$ for the apparent filtered error variance of $\hat{X}_{k|k}$ in (42). The error covariance of $\hat{X}_{k|k}$, $\Sigma_{k|k}$, is given by (3) as:

$$\Sigma_{k|k} = WRW^T$$
$$= [H^T R_{(1+\alpha)}^{-1} H]^{-1} H^T R_{(1+\alpha)}^{-1} R R_{(1+\alpha)}^{-1} H [H^T R_{(1+\alpha)}^{-1} H]^{-1}$$
$$= \Sigma_{(1+\alpha),k|k} \Sigma_{(1+\alpha)^2,k|k}^{-1} \Sigma_{(1+\alpha),k|k} \quad (44)$$

In (44), the inflated filtered error covariance, $\Sigma_{\beta,k|k}$, where $\beta$ denotes the multiplicative inflation factor, is given by:

$$\Sigma_{\beta,k|k} = \beta\Sigma_{k|k-1} - \beta\Sigma_{k|k-1}H_k^T[H_k\beta\Sigma_{k|k-1} + R_k]^{-1}H_k\beta\Sigma_{k|k-1}$$
$$= [H_k^T R_k^{-1} H_k + (\beta\Sigma_{k|k-1})^{-1}]^{-1} \quad (45)$$

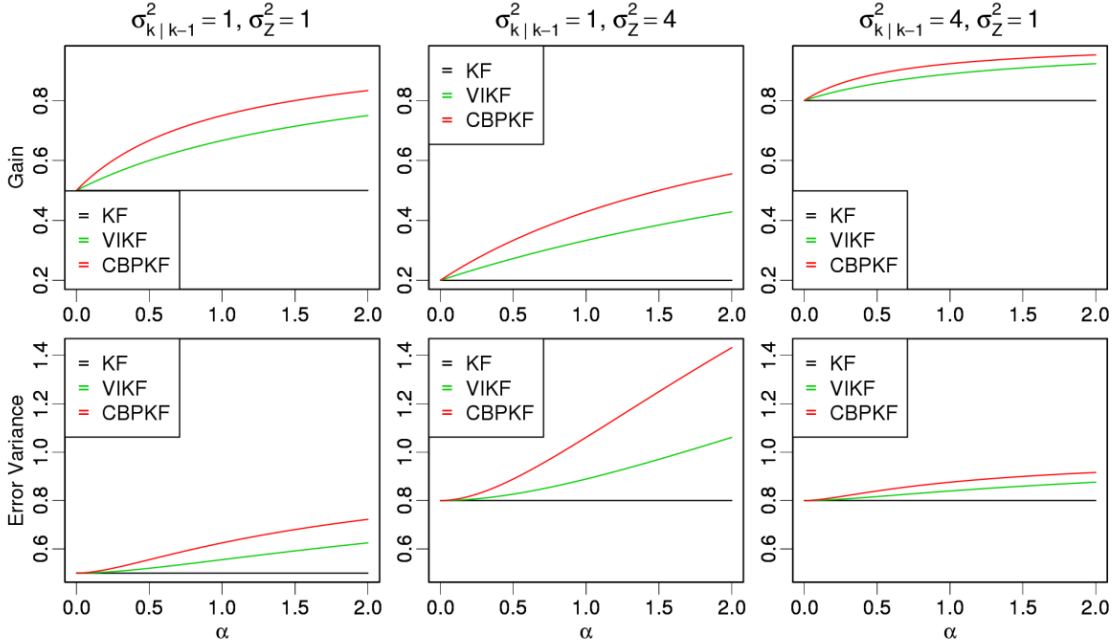

Fig. 1. Comparison of $\kappa_k$ and $\sigma_{k|k}^2$ for KF, the VIKF approximation and CBPKF for three different cases: $\sigma_{k|k-1}^2 = 1$ and $\sigma_Z^2 = 1$ (left), $\sigma_{k|k-1}^2 = 1$ and $\sigma_Z^2 = 4$ (middle), $\sigma_{k|k-1}^2 = 4$ and $\sigma_Z^2 = 1$ (right).

TABLE I
COMPARISON OF GAIN AND FILTERED ERROR VARIANCE AMONG KF, THE VIKF APPROXIMATION, AND CBPKF.

|  | Gain, $\kappa_k$ | Filtered error variance, $\sigma_{k|k}^2$ |
|---|---|---|
| KF | $\dfrac{h\sigma_{k|k-1}^2}{h^2\sigma_{k|k-1}^2 + \sigma_Z^2}$ | $\dfrac{\sigma_Z^2}{h^2\sigma_{k|k-1}^2 + \sigma_Z^2}\sigma_{k|k-1}^2$ |
| VIKF approx. | $\dfrac{h(1+\alpha)\sigma_{k|k-1}^2}{h^2(1+\alpha)\sigma_{k|k-1}^2 + \sigma_Z^2}$ | $\dfrac{\{(1+\alpha)^2 h^2 \sigma_{k|k-1}^2 + \sigma_Z^2\}\sigma_Z^2}{\{(1+\alpha)h^2\sigma_{k|k-1}^2 + \sigma_Z^2\}^2}\sigma_{k|k-1}^2$ |
| CBPKF | $\dfrac{h(1+2\alpha)\sigma_{k|k-1}^2}{h^2(1+2\alpha)\sigma_{k|k-1}^2 + \sigma_Z^2}$ | $\dfrac{\{(1+2\alpha)^2 h^2 \sigma_{k|k-1}^2 + \sigma_Z^2\}\sigma_Z^2}{\{(1+2\alpha)h^2\sigma_{k|k-1}^2 + \sigma_Z^2\}^2}\sigma_{k|k-1}^2$ |

Computationally, evaluation of (43) and (44) requires solving two (m×n) and an (m×m) linear systems. As in the original formulation of CBPKF, iterative reduction of $\alpha$ is necessary to ensure $\Sigma_{k|k} \leq \Sigma_{k|k-1}$. The above approximation assumes that the CB penalty, $\Sigma_{CB}$, is proportional to the error covariance, $\Sigma_{EV}$. To help ascertain how KF, CBPKF and the VIKF approximation may differ, we compare in Table I their analytical solutions for gain $\kappa_k$, and filtered error variance $\sigma_{k|k}^2$ for the 1D case of m=n=1. The table indicates that the VIKF approximation and CBPKF are identical for the 1D problem except that the CB penalty for CBPKF is twice as large as that for the VIKF approximation. To visualize the differences, Fig. 1 shows $\kappa_k$ and $\sigma_{k|k}^2$ for KF, the VIKF approximation and CBPKF for the three cases of $\sigma_{k|k-1}^2 = 1$ and $\sigma_Z^2 = 1$ (left), $\sigma_{k|k-1}^2 = 1$ and $\sigma_Z^2 = 4$ (middle), and $\sigma_{k|k-1}^2 = 4$ and $\sigma_Z^2 = 1$ (right). For all cases, we set h to unity and varied $\alpha$ from 0 to 1. The figure indicates that, compared to the KF solution, the VIKF approximation and the CBPKF solution prescribe appreciably larger gains, that the increase in gain is larger for larger α, and that the CBPKF gain is larger than the gain in the VIKF approximation for the same value of α. The figure also indicates that, compared to KF error variance, CBPKF error variance is larger, and that the increase in error variance is larger for larger α. Note that the differences between the KF and CBPKF solutions are the smallest for $\sigma_{k|k-1}^2 > \sigma_Z^2$, a reflection of the diminished impact of CB owing to the comparatively smaller uncertainty in the observations. The above development suggests that one may be able to approximate CBPKF very closely with the VIKF-based formulation by adjusting α in the latter. Below, we evaluate the performance of CBPKF relative to KF and the VIKF-based approximation of CBPKF.

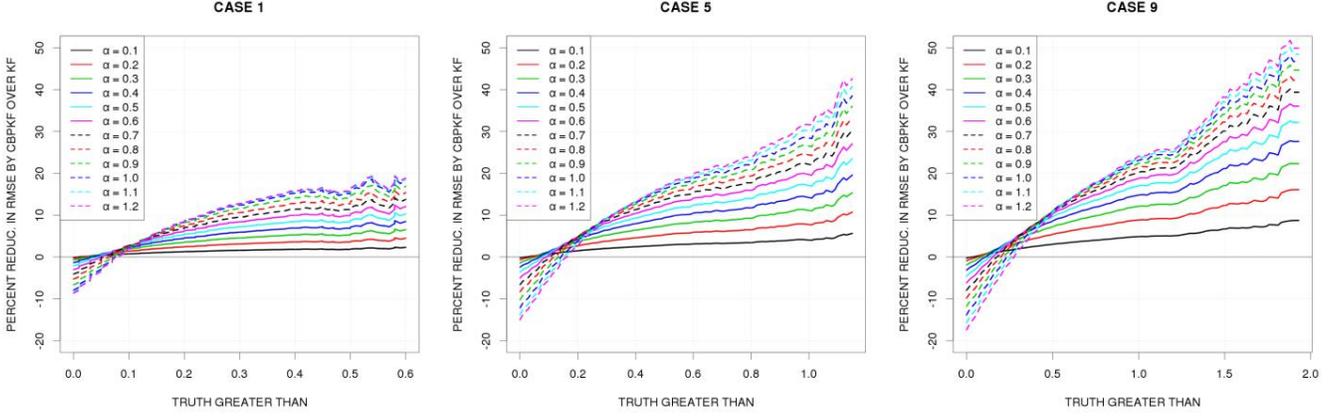

Fig. 2. Percent reduction in RMSE by CBPKF over KF for a range of values of $\alpha$ for Cases 1 (left), 5 (middle) and 9 (right).

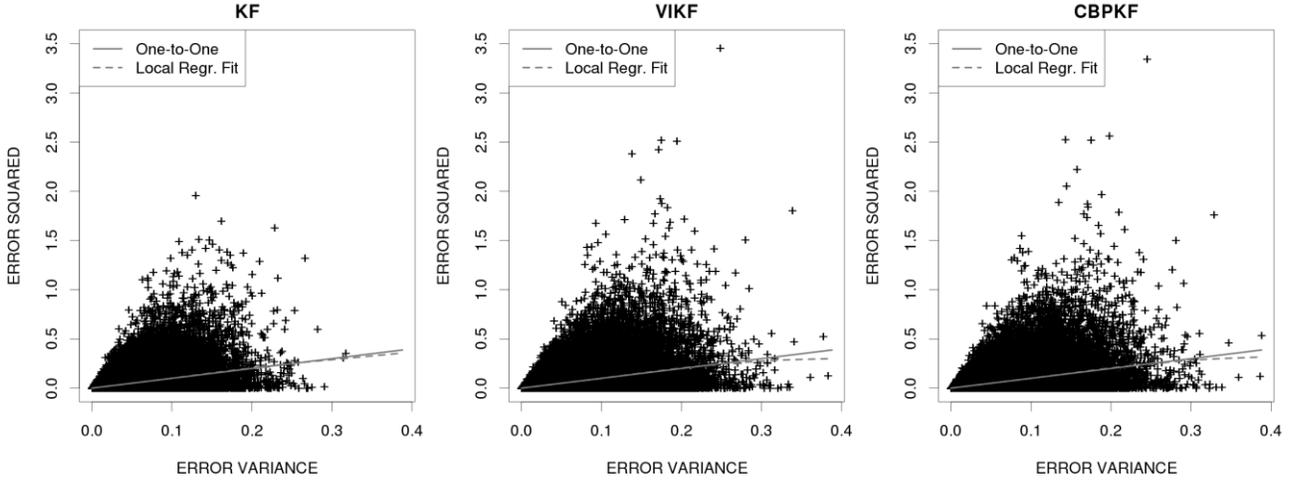

Fig. 3. Filtered error variance vs. error squared for KF (left), the VIKF approximation (middle) and CBPKF (right). The one-to-one line is shown in black and the local regression fit is shown in green.

## V. EVALUATION AND RESULTS

For comparative evaluation, we carried out the synthetic experiments of [24]. We assume the following linear dynamical and observation models with perfectly known statistical parameters:

$$X_k = \Phi_{k-1} X_{k-1} + W_{k-1} \tag{46}$$

$$Z_k = H_k X_k + V_k \tag{47}$$

where $X_k$ and $X_{k-1}$ denote the state vectors at time steps k and k-1, respectively, $\Phi_{k-1}$ denotes the state transition matrix at time step k-1 assumed as $\Phi_{k-1} = \varphi_{k-1} I$, $W_{k-1}$ denotes the white noise vector, $w_{j,k-1} \sim N(0, \sigma_{w_{k-1}}^2)$, j=1,...,m, with $Q_{k-1} = E[W_{k-1} W_{k-1}^T]$, and $V_k$ denotes the observation error vector, $v_{i,k} \sim N(0, \sigma_{v_k}^2)$, i=1,...,n. The number of observations, n, is assumed to be time-invariant. The observation errors are assumed to be independent among themselves and of the true state. To assess comparative performance under widely varying conditions, we randomly perturbed $\varphi_{k-1}$, $\sigma_{w,k-1}$ and $\sigma_{v,k}$ above according to (48) through (50) below, and used only those deviates that satisfy the bounds:

$$\varphi_{k-1}^p = \varphi_{k-1} + \gamma_\varphi \varepsilon_\varphi, \quad 0.5 \leq \varphi_{k-1}^p \leq 0.95 \tag{48}$$

$$\sigma_{w,k-1}^p = \sigma_{w,k-1} + \gamma_w \varepsilon_w, \quad \sigma_{w,k-1}^p \geq 0.01 \tag{49}$$

$$\sigma_{v,k}^p = \sigma_{v,k} + \gamma_v \varepsilon_v, \quad \sigma_{v,k}^p \geq 0.01 \tag{50}$$

In the above, the superscript p signifies that the variable is a perturbation, $\varepsilon_\varphi$, $\varepsilon_w$ and $\varepsilon_v$ denote the normally-distributed white noise for the respective variables, and $\gamma_\varphi$, $\gamma_w$ and $\gamma_v$ denote the standard deviations of the white noise added to $\varphi_{k-1}$, $\sigma_{w,k-1}$ and $\sigma_{v,k}$, respectively. The parameter settings (see Table I) are chosen to encompass less predictable (small $\varphi_{k-1}$) to more predictable (large $\varphi_{k-1}$) processes, certain (small $\sigma_{w,k-1}$) to uncertain (large $\sigma_{w,k-1}$) model dynamics, and more informative (small $\sigma_{v,k}$) to less informative (large $\sigma_{v,k}$) observations. The bounds for $\varphi_{k-1}^p$ in (48) is based on the range of lag-1 serial correlation representing moderate to high predictability where CBPKF and KF are likely to differ the most. The bounding of the perturbed values $\sigma_{w,k-1}^p$ and $\sigma_{v,k}^p$ in (49) and (50), respectively, is necessary to avoid the observational or model prediction uncertainty becoming unrealistically small. Very small $\sigma_{w,k-1}^p$ and $\sigma_{v,k}^p$ render the information content of the model prediction, $\Sigma_{k|k-1}$, and the observation, $Z_k$, respectively, very large, and hence keep the filters operating in unrealistically favorable conditions for extended periods of time. We then apply KF, CBPKF and the VIKF approximation to obtain $\hat{X}_{k|k}$ and $\Sigma_{k|k}$, and verify them against the assumed truth. To evaluate the performance of CBPKF relative to KF, we calculate percent reduction in root mean square error (RMSE) by CBPKF over KF conditional on the true state exceeding some threshold between 0 and the largest truth.

TABLE II
PARAMETER SETTINGS FOR THE 12 CASES CONSIDERED.

| Group | Case | $\sigma_{w,k-1}$ | $\gamma_w$ | $\sigma_{v,k}$ | $\gamma_v$ | $\varphi_{k-1}$ | $\gamma_\varphi$ |
|---|---|---|---|---|---|---|---|
| 1 | 1 | 0.1 | 0.01 | 1.5 | 0.4 | 0.7 | 0.1 |
|   | 2 | 0.1 | 0.01 | 1.5 | 0.4 | 0.7 | 0.8 |
|   | 3 | 0.1 | 0.01 | 1.5 | 1.2 | 0.7 | 0.1 |
|   | 4 | 0.1 | 0.01 | 1.5 | 1.2 | 0.7 | 0.8 |
| 2 | 5 | 0.1 | 0.1 | 1.5 | 0.4 | 0.7 | 0.1 |
|   | 6 | 0.1 | 0.1 | 1.5 | 0.4 | 0.7 | 0.8 |
|   | 7 | 0.1 | 0.1 | 1.5 | 1.2 | 0.7 | 0.1 |
|   | 8 | 0.1 | 0.1 | 1.5 | 1.2 | 0.7 | 0.8 |
| 3 | 9 | 0.1 | 0.2 | 1.5 | 0.4 | 0.7 | 0.1 |
|   | 10 | 0.1 | 0.2 | 1.5 | 0.4 | 0.7 | 0.8 |
|   | 11 | 0.1 | 0.2 | 1.5 | 1.2 | 0.7 | 0.1 |
|   | 12 | 0.1 | 0.2 | 1.5 | 1.2 | 0.7 | 0.8 |

TABLE III
COMPARISON OF COMPUTING TIME AMONG KF, CBPKF AND VIKF APPROXIMATION.

| Dimensionality | | Normalized computing time | | |
|---|---|---|---|---|
| m | n | KF | CBPKF | VIKF approx. |
| 1 | 10 | 1 | 5.23 | 1.51 |
| 1 | 40 | 1 | 18.41 | 2.74 |
| 5 | 10 | 1 | 6.44 | 1.67 |
| 5 | 40 | 1 | 24.03 | 2.88 |
| 10 | 10 | 1 | 14.27 | 2.03 |
| 10 | 40 | 1 | 27.96 | 3.46 |

Fig. 2 show the percent reduction in RMSE by CBPKF over KF for Cases 1 (left), 5 (middle) and 9 (right) representing Groups 1, 2 and 3 in Table I, respectively. The three groups differ most significantly in the variability of the dynamical model error, $\gamma_w$, and may be characterized as nearly stationary (Group 1), nonstationary (Group 2), and highly nonstationary (Group 3). The range of $\alpha$ values used is [0.1, 1.2] with an increment of 0.1. The numbers of state variables, observations, and updating cycles used in Fig. 2 are 1, 10, and 100,000 for all cases. The dotted line at 10% reduction in the figure serves as a reference for significant improvement. The figure shows that, at the extreme end of the tail, CBPKF with $\alpha$ of 0.7, 0.6 and 0.5 reduces RMSE by about 15, 25 and 30% for Cases 1, 5 and 9, respectively, but at the expense of increasing unconditional RMSE by about 5%. The general pattern of reduction in RMSE for other cases in Table I is similar within each group and is not shown. We only note here that larger variability in observational uncertainty (i.e., larger $\gamma_v$) reduces the relative performance of CBPKF somewhat, and that the magnitude of variability in predictability (i.e., $\gamma_\varphi$) has relatively small impact on the relative performance.

It was seen in Table I that the VIKF approximation is identical to CBPKF for m=n=1 but for the multiplicative scaler weight for the CB penalty. Numerical experiments indicate that, whereas the above relationship does not hold for other m or n, one may very closely approximate CBPKF with the VIKF-based formulation by adjusting $\alpha$. For example, the VIKF approximation with $\alpha$ increased by a factor of 1.25 to 1.90 differ from CBPKF only by 1% or less for all 12 cases in Table II with m=1 and n=10. The above findings indicate that the VIKF approximation may be used as a computationally less expensive alternative for CBPKF. Table III compares the CPU time among KF, CBPKF and the VIKF approximation for 6 different combinations of m and n based using Intel(R) Xeon(R) Gold 6152 CPU @ 2.10GHz. The computing time is reported in multiples of the KF's. Note that the original formulation of CBPKF quickly becomes extremely expensive as the dimensionality of the problem increases whereas the CPU time of the VIKF approximation stays under 3.5 times that of KF for the size of the problems considered.

If the filtered error variance is unbiased, one would expect the mean of the actual error squared associated with the variance to be approximately the same as the variance itself. To verify this, we show in Fig. 3 the filtered error variance vs. the actual error squared for KF (left), the VIKF approximation (middle) and CBPKF (right) for all ranges of filtered error variance. For reference, we plot the one-to-one line representing the unbiased error variance conditional on the magnitude of the filtered error variance and overlay the local regression fit through the actual data points using the R package locfit[34]. The figure shows that all three provide conditionally unbiased estimates of filtered error variance as theoretically expected, and that the VIKF approximation and CBPKF results are extremely similar to each other.

## VI. ADAPTIVE CBPKF

Whereas CBPKF or the VIKF approximation significantly improves the accuracy of the estimates over the tails, it deteriorates performance near the median. Fig. 2 suggests that, if $\alpha$ can be prescribed adaptively such that a small/large CB penalty is effected when the system is in the normal/extreme state, the unconditional performance of CBPKF would improve. Because the true state of the system is not known, adaptively specifying $\alpha$ is necessarily an uncertain proposition. There are, however, certain applications in which the normal-vs.-extreme state of the system may be ascertained with higher accuracy than others. For example, the soil moisture state of a catchment may be estimated from assimilating precipitation and streamflow data into hydrologic models[35][36][37][38][39][40]. If $\alpha$ is prescribed adaptively based on the best available estimate of the state of the catchment, one may expect improved performance in hydrologic forecasting. In this section, we apply adaptive CBPKF in the synthetic experiment and assess its performance. An obvious strategy for adaptively filtering is to parameterize $\alpha$ in terms of the KF estimate (i.e., the CBPKF estimate with $\alpha = 0$) as the best guess for the true state. The premise of this strategy is that, though it may be conditionally biased, the KF estimate fuses the information available from both the observations and the dynamical model, and hence best captures the relationship between $\alpha$ and the departure of the state of the system from median. A similar approach has been used in fusing radar rainfall data and rain gauge observations for multisensor precipitation estimation in which ordinary cokriging estimate was used to prescribe $\alpha$ in CB-penalized cokriging[29].

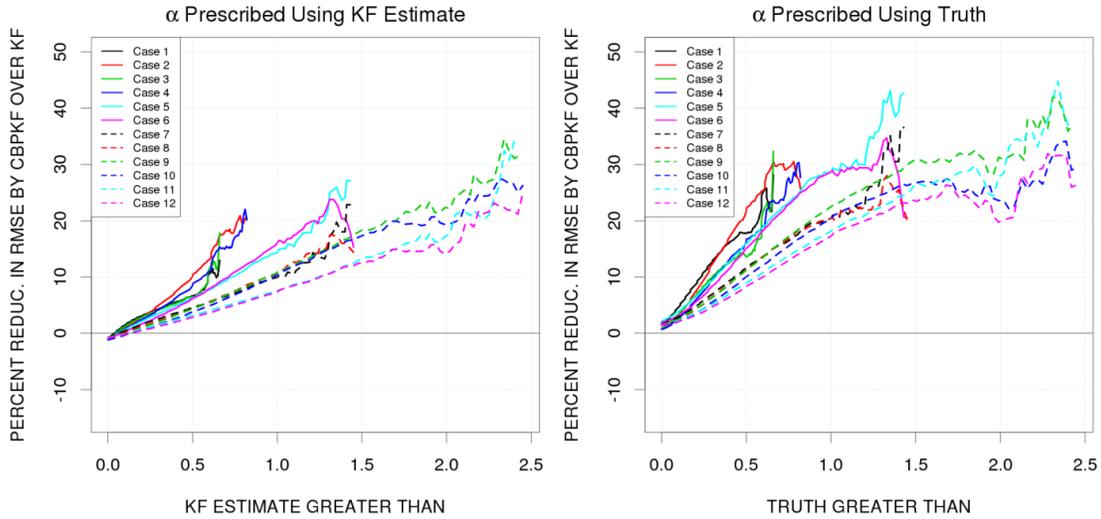

Fig. 4. Percent reduction in RMSE by adaptive CBPKF over KF in which $\alpha$ is prescribed using the KF estimate (left) and the truth (right).

Necessarily, the effectiveness of the above strategy depends on the skill of the KF estimate; if the skill is very low, one may not expect significant improvement. Fig. 2 suggests that, qualitatively, α should increase as the state becomes more extreme. To that end, we employed the following model for time-varying $\alpha$:

$$\alpha_k = \gamma \|\hat{X}_{k|k}^{KF}\| \tag{51}$$

where $\alpha_k$ denotes the multiplicative CB penalty factor for CBPKF at time step k, $\|\hat{X}_{k|k}^{KF}\|$ denotes some norm of the KF estimate at time step k, and $\gamma$ denotes the proportionality constant.

Fig. 4a shows the RMSE reduction by adaptive CBPKF over KF with $\alpha_k = \gamma|\hat{X}_{k|k}^{KF}|$ for the 12 cases in Table II m=1 and n=10. The $\gamma$ values used were 3.0, 1.0 and 0.5 for Groups 1, 2 and 3 in Table II, respectively. The figure shows that adaptive CBPKF performs comparably to KF in the unconditional sense while substantially improving performance in the tails. The rate of reduction in RMSE with respect to the increasing conditioning truth, however, is now slower than that seen in Fig. 2 due to the occurrences of incorrectly specified α. To assess the uppermost bound of the feasible performance of adaptive

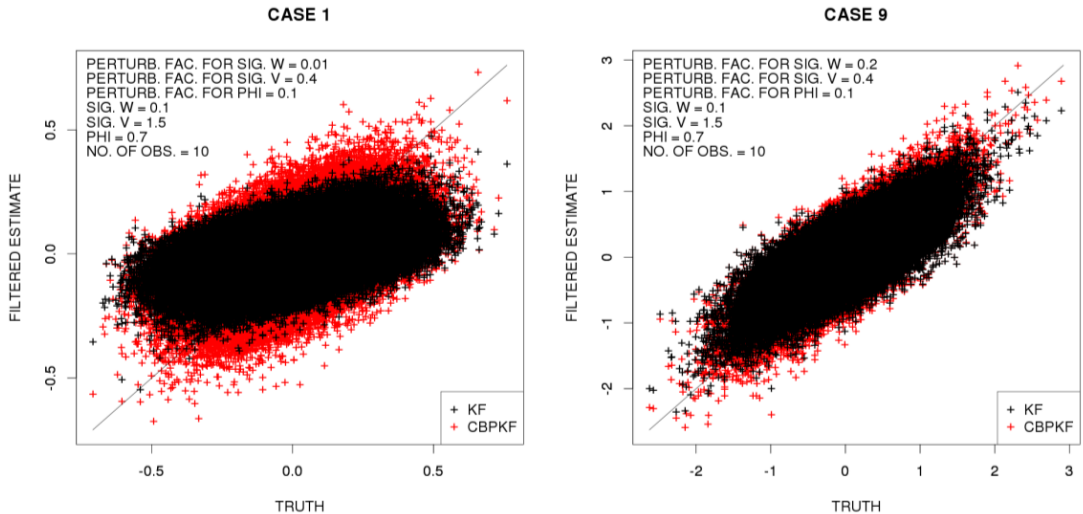

Fig. 5. Example scatter plots of KF (black) and adaptive CBPKF (red) estimates vs. truth for Cases 1 (left) and 9 (right) in Table II.

CBPKF, we also specified $\alpha$ with perfect accuracy under (51) via $\alpha_k = \gamma |X_k|$ where $X_k$ denotes the true state. The results are shown in Fig. 4b for which the $\gamma$ values used were 3.0, 1.5 and 1.0 for Groups 1, 2 and 3 in Table II, respectively. The figure indicates that adaptive CBPKF with perfectly prescribed $\alpha$ greatly improves performance, outperforming KF even in the unconditional sense. Fig. 4 suggests that, if $\alpha$ can be prescribed more accurately with additional sources of information, the performance of adaptive CBPKF may be improved beyond the level seen in Fig. 4a. Finally, we show in Fig. 5 the example scatter plots of the KF (black) and adaptive CBPKF (red) estimates vs. truth. They are for Cases 1 and 9 in Table II representing Groups 1 and 3, respectively. It is readily seen that the CBPKF significantly reduces CB in the tails while keeping its estimates close to the KF estimates in normal ranges.

## VII. Conclusions

Conditional bias-penalized Kalman filter (CBPKF) has recently been developed to improve estimation and prediction of extremes. The original formulation, however, is computationally very expensive, and deteriorates performance in the normal ranges relative to KF. In this work, we present a computationally less expensive alternative based on the variance-inflated KF (VIKF) approximation, and improve unconditional performance by adaptively prescribing the weight for the CB penalty. For evaluation, we carried out synthetic experiments using linear systems with varying degrees of dynamical model uncertainty, observational uncertainty, and predictability. The results indicate that the VIKF-based approximation of CBPKF provides a computationally much less expensive alternative to the original formulation, and that adaptive CBPKF performs comparably to KF in the unconditional sense while improving estimation of extremes by about 20 to 30% over KF. It is also shown that additional improvement may be possible by improving adaptive prescription of the weight to the CB penalty using additional sources of information. The findings indicate that adaptive CBPKF offers a significant addition to the dynamic filtering methods for general application in signal processing and, in particular, when or where estimation of extremes is of importance. The findings in this work are based on idealized synthetic experiments that satisfy linearity and normality. Additional research is needed to assess performance for non-normal problems and for nonlinear problems using the ensemble extension[30], and to prescribe the weight for the CB penalty more skillfully.